# Dynamic Effective Medium Theory for Two-Dimensional Non-Magnetic Metamaterial Lattices using Multipole Expansion

Ioannis Chremmos[1], Efthymios Kallos[2], Melpomeni Giamalaki[3],
Vassilios Yannopapas[4], Emmanuel Paspalakis[2]

[1]Max-Planck-Institute for the Science of Light, D-91058, Erlangen, Germany
[2]Department of Materials Science, University of Patras, GR-26504, Patras, Greece
[3]School of Electrical and Computer Engineering, National Technical University, GR 157-73, Athens, Greece
[4]Department of Physics, National Technical University of Athens, GR-15780, Athens, Greece

We present a formulation for deriving effective medium properties of infinitely periodic two-dimensional metamaterial lattice structures beyond the conventional static and quasi-static limits. We utilize the multipole expansions, where the polarization currents associated with the supported Bloch modes are expressed via the electric dipole, magnetic dipole, and electric quadrupole moments per unit length. We then propose a method to calculate the Bloch modes based on the lattice geometry and individual unit element structure. The results revert to well-known formulas in the traditional quasistatic limit and are useful for the homogenization of nanorod-type metamaterials which are frequently used in optical applications.

*PACS numbers:* 42.25.Fx, 78.67.Pt, 78.67.Qa

## I. INTRODUCTION

Artificial structures consisting of arrays of subwavelength elements are of great interest as they allow unconventional functions such as cloaking,[1-3] optical magnetism,[4, 5] and solar energy applications[6-8]. Such composite structures are often conveniently modelled as equivalent homogeneous media with corresponding effective electromagnetic parameters (effective permittivity and effective permeability). This interpretation, although not always applicable or accurate, significantly simplifies the understanding of the electromagnetic performance of the structure under consideration.

A wide range of homogenization techniques is currently available in the literature, including: the use of simulated or experimentally retrieved scattering parameters (Nicolson-Ross-Weir method)[9-12]; the averaging of simulated fields known inside a simulation domain[13-16]; and the extension of Floquet-based theories with local homogenization schemes[17, 18]. There has also been significant body of work to obtain the so-called *mixing formulas*, which attempt to describe analytically in closed-form expressions the effective parameters of a composite medium[19, 20]. These formulas typically require the use of the (dipole) polarizabilities (electric and/or magnetic) of the unit element of the structure, which is analytically described for regular shapes such as spheres or cylinders.

Most of the above-mentioned techniques are focused on three-dimensional (3D) structures, which can have isotropic responses. While this is desirable in many applications, there is growing interest in two-dimensional (2D) structures, which can have inherently anisotropic properties that can be in principle tuned independently (e.g. for transformation optics applications)[21]. Such rod-type structures, which extend much longer along one dimension in space, have been investigated and experimentally tested in the past, particularly while attempting to generate artificial magnetism [22-25].

Obtaining effective medium properties for 2D lattice structures has been more challenging than their 3D counterparts, mainly due to the inherent anisotropic nature of the structures. While a few attempts have been made to provide an analytical framework for the effective medium properties of 2D metamaterials [26, 27], they are typically limited to the long-wavelength (quasistatic) limit [28, 29], where the wavelength of light is much longer that the inclusion size, or assume only dipole interactions between the inclusions in the lattice[30-32], restricting the accuracy of the methods. Due to the anisotropic nature of the rod geometry, extending beyond the dipole interactions has been particularly challenging because the multipole expansion process is more complicated than the usual 3D expansion that is readily available [33-35].

In this work, we develop a effective medium theory for 2D metamaterial lattices using multipole expansion that extends beyond the quasi-static limit. The theoretical process evolves in two phases. In the first phase, we reformulate Maxwell's equations in terms of the periodic envelopes of Bloch modes in a periodic lattice [36, 37]. In this context and taking into account the subwavelength size of the dielectric elements, the retardation (or propagation) effect of the Bloch mode through a lattice element can be approximated by a Taylor series up to the linear term. This decomposes the effect of the polarization current −the source of the electromagnetic (EM) fields in the lattice− into its electric dipole, magnetic dipole and electric quadrupole moments, which can be treated as independent sources of radiation. The reformulated equations are subsequently averaged over a unit cell to obtain bulk-type propagation equations between the average EM fields and the volume (per unit length) densities of the multipole moments. With a judicious rearrangement, the equations become equivalent to a homogeneous magnetic medium whose permittivity and permeability are identified as the sought effective parameters of our all-dielectric lattice. We derive closed-form expressions for these effective parameters in terms of the densities of the multipole moments. The latter are quantities that have to be derived from the field distribution of the Bloch modes, the computation of which is the focus of the second phase of the analysis.

To determine the Bloch modes of the lattice, we employ a classical method that is very efficient for cylindrical elements and dates back to Rayleigh. In this approach, the transverse field (electric for transverse magnetic and magnetic for transverse electric modes) in the neighborhood of a cylinder is split into a standing regular and an outward radiated (or scattered) singular part. The former is the superposition of the corresponding singular waves radiated by the other elements in the lattice. Using (Graf's) addition theorems for cylindrical wave functions these radiated waves are translated to the local system of cylindrical coordinates and boundary conditions are applied. We end up with a homogeneous linear system for the amplitudes of the scattered field harmonics. The nontrivial solutions of this system yield the momenta of the Bloch modes while the corresponding amplitudes determine the corresponding field distribution. Finally, using formulas that we have previously derived on the multipole expansion of the fields radiated by dielectric cylinders, [38] the amplitudes of the radiated waves can be directly related to the moments of the polarization currents that feed the equations obtained in the first phase of our analysis.

The matrix of the system obtained with Rayleigh's method, contains lattice sums of cylindrical wave functions over the infinite periodic lattice. The lattice sums determine the effect or interaction between a field harmonic in an element and a field harmonic in all other elements in the lattice, with the Bloch momentum acting as a parameter. An efficient method has to be adopted for computing these sums since a direct summation is futile due to their conditional convergence. We here opt for a method that utilizes two equivalent representations of the quasi-periodic Green's function in the direct and reciprocal lattice spaces and is described in detail in Appendix A.

In this paper, we present the analytical formulation of the proposed method, showcasing the simpler scenarios and validating them with numerical calculations. A more extensive numerical and full wave simulation validation campaign is being planned for a subsequent paper. The rest of this paper is organized as follow. Section II presents the derivation of the effective medium formulas for an infinite 2D lattice, as a function of the Bloch modes and the lattice and inclusion geometry. In Section III we calculate the required multipole moments for a given Bloch mode supported by the lattice. Finally, Section IV presents numerical examples that illustrate the performance of our method compared to the simpler quasistatic methods. The paper also includes three appendices: Appendix A deals with the efficient calculation of the lattice sums that appear in the method. Appendix B provides details on the calculation of the effective parameter formulas. Appendix C demonstrates that the dynamic formulas obtained with this method resort to the well-known quasistatic ones in the long-wavelength limit.

## II. DERIVATION OF THE EFFECTIVE MEDIUM FORMULAS

In this section we derive the effective medium formulas for an infinite 2D periodic lattice of non-magnetic rods. The derivation provides the effective medium response (effective permittivity and permeability) as a function of the lattice geometry and the supported Bloch modes. In the next section we propose a method to calculate the Bloch modes based on a multipole description of the EM response of the system.

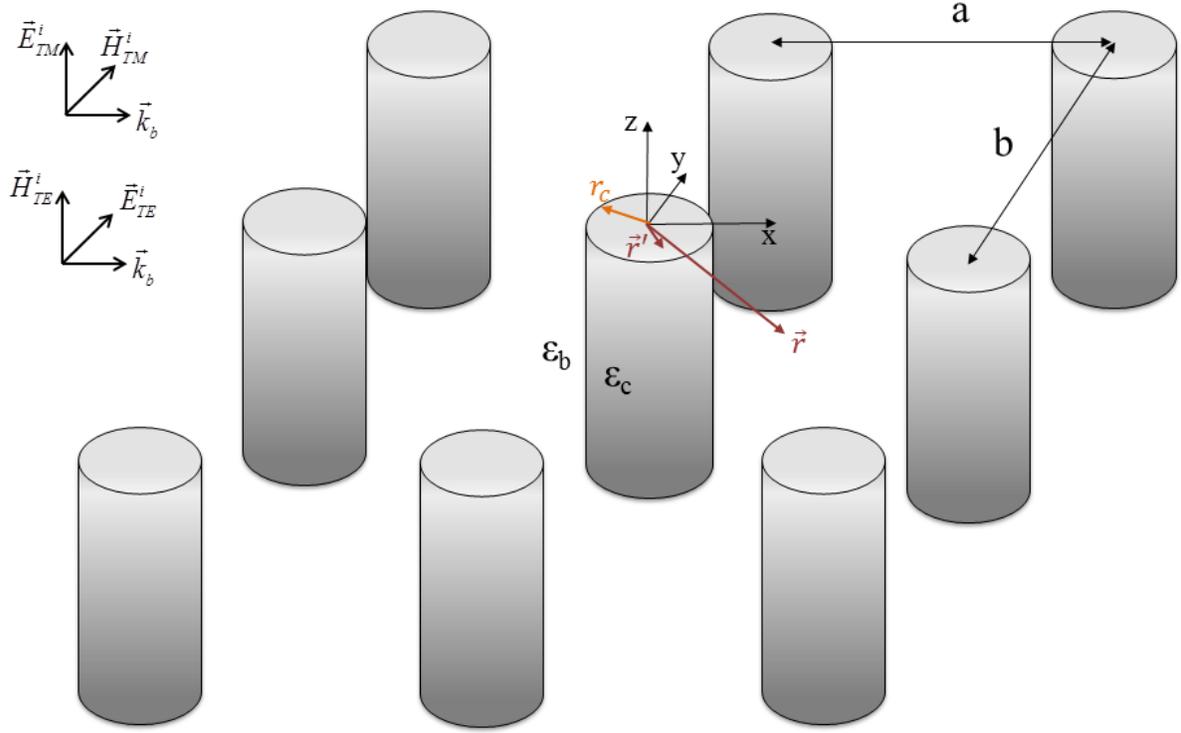

**Figure 1: A section of the infinite 2D lattice of non-magnetic infinitely long cylindrical rods embedded in a host medium.**

We consider an infinite rectangular lattice of parallel, infinitely long, circularly symmetric non-magnetic rods with relative permittivity profile $\varepsilon_c(\vec{r})$ embedded in a homogeneous medium with homogeneous relative permittivity $\varepsilon_b$. Cartesian coordinates are chosen so that $z$ is the direction parallel to the cylinders and the primitive lattice vectors are $a\hat{x}$ and $b\hat{y}$ (see Figure 1). Due to its invariance along $z$ and periodicity along $x$ and $y$, the structure supports Bloch modes the EM field of which can be expressed as

$$\vec{E}(\vec{r}) = \vec{e}(\vec{r}) e^{j\vec{K}\cdot\vec{r}}, \quad \vec{H}(\vec{r}) = \vec{h}(\vec{r}) e^{j\vec{K}\cdot\vec{r}}, \tag{1}$$

where $\vec{K}$ is the Bloch wave vector, and $\vec{e}(\vec{r}), \vec{h}(\vec{r})$ are periodic envelopes that conform to the periodicity of the lattice, i.e. $\vec{e}(\vec{r}+\vec{R}_\ell) = \vec{e}(\vec{r})$, $\vec{h}(\vec{r}+\vec{R}_\ell) = \vec{h}(\vec{r})$, where $\vec{R}_\ell = \hat{x}\ell_1 a + \hat{y}\ell_2 b$ is an arbitrary lattice vector indexed by the pair of integers $\ell = (\ell_1, \ell_2)$. We restrict ourselves to in-plane propagating waves with $K_z = 0$ which allows the decomposition of the EM field into independent Transverse-Magnetic (TM, $H_z = 0$) and Transverse-Electric (TE, $E_z = 0$) states of polarization. Substituting (1) to Maxwell's equations we obtain

$$\nabla \times \vec{e} + j\vec{K} \times \vec{e} = j\omega\mu_0 \vec{h}$$
$$\nabla \times \vec{h} + j\vec{K} \times \vec{h} = -j\omega\varepsilon_0\varepsilon_b \vec{e} + \vec{J}e^{-j\vec{K}\cdot\vec{r}} \qquad (2)$$

where $\vec{J}(\vec{r}) = -j\omega\varepsilon_0(\varepsilon(\vec{r}) - \varepsilon_b)\vec{E}(\vec{r})$ is the polarization current in the rods. Since that polarization current is only supported in the region occupied by the scatterer, one may assume that $\vec{K}\cdot\vec{r} \ll 1$ and approximate

$$\vec{J}e^{-j\vec{K}\cdot\vec{r}} \approx \vec{J}(1 - j\vec{r}\cdot\vec{K}) = \vec{J} - \frac{j}{2}(\vec{J}\otimes\vec{r} + \vec{r}\otimes\vec{J})\cdot\vec{K} + \frac{j}{2}\vec{K}\times(\vec{r}\times\vec{J}) \qquad (3)$$

where $\otimes$ denotes the tensor product of corresponding vectors. The above equation approximates the retardation (or propagation) effect of the Bloch mode, which is expressed through the exponential phase factor, in terms of its two first Taylor terms. Substituting (3) in (2) and integrating over the rectangular unit cell area we obtain

$$\vec{K} \times \vec{E}_{av} = \omega\mu_0 \vec{H}_{av}$$
$$\vec{K} \times \vec{H}_{av} \cong -\omega\varepsilon_0\varepsilon_b \vec{E}_{av} - \omega\vec{P}_{av} + \frac{j\omega}{2}\ddot{Q}_{av}\cdot\vec{K} + \vec{K}\times\vec{M}_{av} \qquad (4)$$

where the average volume densities of electric dipole moment $\vec{P}_{av}$, electric quadrupole moment $\ddot{Q}_{av}$, and magnetic dipole moment $\vec{M}_{av}$ (per unit length) of the polarization current appear as defined via [39]

$$\vec{P}_{av} = \frac{\vec{p}}{S_{cell}} = \frac{1}{S_{cell}} \cdot \frac{j}{\omega} \int_{u.c.} \vec{J}(\vec{r}')dS'$$
$$\vec{M}_{av} = \frac{\vec{m}}{S_{cell}} = \frac{1}{S_{cell}} \cdot \frac{1}{2} \int_{u.c.} \vec{r}' \times \vec{J}(\vec{r}')dS' \qquad (5)$$
$$\ddot{Q}_{av} = \frac{\ddot{Q}}{S_{cell}} = \frac{1}{S_{cell}} \cdot \frac{j}{\omega} \int_{u.c.} (\vec{J}\otimes\vec{r}' + \vec{r}'\otimes\vec{J})dS'$$

In the latter equation the integration takes place over the area of the unit cell (u.c.), which is equal to $S_{cell} = a \cdot b$, and $\vec{p}, \vec{m}, \ddot{Q}$ are the multipole moments per unit length. Similarly, we have defined the average electric and magnetic fields over a unit cell as

$$\begin{pmatrix} \vec{E}_{av} \\ \vec{H}_{av} \end{pmatrix} = \frac{1}{S_{cell}} \int_{S_{cell}} \begin{pmatrix} \vec{e}(\vec{r}') \\ \vec{h}(\vec{r}') \end{pmatrix} dS' \tag{6}$$

After some algebraic manipulations, Eq. (4) can be solved for the average field quantities as follows

$$\vec{E}_{av} = \frac{\left(k_b^2 \ddot{I} - \vec{K}\vec{K}\right)}{\omega \varepsilon_0 \varepsilon_b \left(k_b^2 - |\vec{K}|^2\right)} \cdot \left(-\omega \vec{P}_{av} + \frac{j\omega}{2} \ddot{Q}_{av} \cdot \vec{K} + \vec{K} \times \vec{M}_{av}\right)$$

$$\vec{H}_{av} = \frac{\vec{K}}{k_b^2 - |\vec{K}|^2} \times \left(-\omega \vec{P}_{av} + \frac{j\omega}{2} \ddot{Q}_{av} \cdot \vec{K} + \vec{K} \times \vec{M}_{av}\right) \tag{7}$$

Here $k_b = \omega / c_b$ is the wavenumber in the background medium with $c_b = c / \sqrt{\varepsilon_b}$ being the speed of light in that medium. The latter equations relate the sources (polarization currents expressed via the average volume densities of their multipole moments) with the produced averaged fields over a unit cell for a given Bloch mode in the lattice. It is worth noting that in the case of TM waves, the current $\vec{J}$ flows along the $z$ axis ($\vec{K} \cdot \vec{J} = 0$) and one can show that $j\omega \ddot{Q}_{av} \cdot \vec{K} = 2\vec{K} \times \vec{M}_{av}$ [38], implying that the quadrupole moment degenerates to a magnetic dipole one. This does not occur in the TE case or in 3D structures.

In order to obtain the effective parameters of the medium, we observe that Eqs. (4) can be rewritten as

$$\vec{K} \times \vec{E}_{eq} = \omega \vec{B}_{eq}$$
$$\vec{K} \times \vec{H}_{eq} \cong -\omega \vec{D}_{eq}, \tag{8}$$

where we introduce the equivalent field quantities through

$$\vec{E}_{eq} = \vec{E}_{av} - \frac{j}{2\varepsilon_0 \varepsilon_b} \ddot{Q}_{av} \cdot \vec{K}$$

$$\vec{B}_{eq} = \mu_0 \vec{H}_{av} - \frac{j}{2\omega\varepsilon_0 \varepsilon_b} \vec{K} \times \ddot{Q}_{av} \cdot \vec{K}$$

$$\vec{H}_{eq} = \vec{H}_{av} - \vec{M}_{av}$$

$$\vec{D}_{eq} = \varepsilon_0 \varepsilon_b \vec{E}_{av} + \vec{P}_{av} - \frac{j}{2} \ddot{Q}_{av} \cdot \vec{K}$$

(9)

The above equations provide physical insight on how artificial magnetism can arise from non-magnetic materials. The equations describe an *equivalent* homogeneous medium that supports a plane wave with wavevector $\vec{K}$, equal to the momentum of Bloch waves inside the metamaterial. However, while the original lattice is nonmagnetic, its homogeneous equivalent is not, as follows by the fact $\vec{B}_{eq} \neq \mu_0 \vec{H}_{eq}$ which is easily seen from the second and third equations in (9). The apparent magnetic behavior of the equivalent medium arises from the *electric* polarization currents associated with the Bloch mode in the lattice, which give rise to an equivalent magnetic response. Thus, the system can be viewed either as the original non-magnetic medium with a magnetic response arising from the supported electric polarization currents only, or as the equivalent magnetic medium with the magnetic response arising from an effective permeability $\ddot{\mu}_{eff}$ which will be defined shortly.

Also, recall that in the TM case the above definitions simplify by dropping the quadrupole terms and doubling the magnitude of vector $\vec{M}_{av}$. Moreover, in this case it is easy to see that $\vec{E}_{eq} = \vec{E}_{av}$ and $\vec{B}_{eq} = \mu_0 \vec{H}_{av} = \vec{B}_{av}$, namely the equivalent medium supports the same fundamental fields $\vec{E}, \vec{B}$ with the non-magnetic metamaterial.

We now write the constitutive relations of the homogeneous medium to get

$$\vec{D}_{eq} = \varepsilon_0 \ddot{\varepsilon}_{eff} \cdot \vec{E}_{eq}$$

$$\vec{B}_{eq} = \mu_0 \ddot{\mu}_{eff} \cdot \vec{H}_{eq}$$

(10)

where $\ddot{\varepsilon}_{eff}, \ddot{\mu}_{eff}$ are the tensors of relative effective permittivity and permeability of our system. From Eqs. (10) and using (7) and (8), we can solve for the various elements of the permittivity and permeability tensors. In the case of the $zz$ components of the tensors, these are

$$\frac{\varepsilon_{eff}^{zz}}{\varepsilon_b} = 1 + \left(\left|\frac{\vec{K}}{k_b}\right|^2 - 1\right)\frac{P_{z,av}k_b c_b}{P_{z,av}k_b c_b - 2\hat{z}\cdot(\vec{K}\times\vec{M}_{av})}$$

$$\mu_{eff}^{zz} = 1 + \left(\left|\frac{\vec{K}}{k_b}\right|^2 - 1\right)\frac{k_b^2 M_{z,av} - (j\omega/2)\hat{z}\cdot\left[\vec{K}\times(\ddot{Q}_{av}\cdot\vec{K})\right]}{k_b^2 M_{z,av} + \hat{z}\cdot(\vec{K}\times\vec{P}_{av}k_b c_b) - (j\omega/2)\hat{z}\cdot\left[\vec{K}\times(\ddot{Q}_{av}\cdot\vec{K})\right]}$$

(11)

where the various vector products are written explicitly

$$\vec{K}\times\vec{P}_{av} = \hat{z}(K_x P_{y,av} - K_y P_{x,av})$$
$$\vec{K}\times(\ddot{Q}_{av}\cdot\vec{K}) = \hat{z}\left[K_x K_y(Q_{yy,av} - Q_{xx,av}) + (K_x^2 - K_y^2)Q_{xy,av}\right]$$
$$\vec{K}\times\vec{M}_{av} = \hat{z}(K_x M_{y,av} - K_y M_{x,av})$$

(12)

Here the indices in the average polarization densities indicate the corresponding vector or tensor component. The derivation of the other entries of the tensor elements is more involved. Their corresponding formulas result from a 3×3 linear system and are too cumbersome to be quoted here. Their derivation is discussed in Appendix B.

We have thus expressed the effective parameters of the nanorod medium as closed form expressions involving the *x*, *y* components of the Bloch momentum and the moments of the electric polarization current of the Bloch mode supported by the lattice at a given frequency.

### III. EVALUATION OF THE MULTIPOLE MOMENTS FOR A BLOCH MODE

In this section we describe how the needed information for the Bloch modes can be obtained for the evaluation of the effective parameters of our system. The first part of our approach is based on Rayleigh's method [40, 41] which provides an elegant analytical formulation of the multiscattering process in the 2D lattice. In this context, the field function in the vicinity of an arbitrary cylinder, say the one centered at $\vec{R}_{(0,0)} = \vec{0}$, is expanded in terms of separable solutions of the Helmholtz equation, in particular the regular and outgoing cylindrical wavefunctions

$$\hat{\psi}_m(\vec{r}) = J_m(k_b r)e^{jm\varphi}$$
$$\psi_m(\vec{r}) = H_m^{(1)}(k_b r)e^{jm\varphi}$$

(13)

where $J_m$ and $H_m^{(1)}$ are, respectively, the Bessel and Hankel functions (of the first kind) with integer order $m$ and $(r,\varphi,z)$ are the cylindrical coordinates of field point $\vec{r}$. The $z$ electric or magnetic field component, noted as $u(\vec{r})$, can then be expanded as

$$u(\vec{r}) = \sum_{m=-\infty}^{+\infty} \left( A_m \hat{\psi}_m(\vec{r}) + B_m \psi_m(\vec{r}) \right) \tag{14}$$

which is valid inside the "source-free" annulus $r_c < r < \min(a,b) - r_c$. The coefficients $A_m$ and $B_m$ are the amplitudes of the regular and scattered wavefunctions, related via $A_m = T_{|m|}^{-1} B_m$, where $T_{|m|}$ are the well-known cylindrical Mie scattering coefficients given by

$$T_m = \left( -1 - j \frac{Y'_m(k_b r_c) J_m(k_c r_c) - (\varepsilon_c/\varepsilon_b)^{\pm 1/2} Y_m(k_b r_c) J'_m(k_c r_c)}{J'_m(x_b) J_m(k_b r_c) - (\varepsilon_c/\varepsilon_b)^{\pm 1/2} J_m(k_b r_c) J'_m(k_c r_c)} \right)^{-1}$$

where $\pm$ applies to the TM/TE case [38, 42]. Note that the sign of $m$ is immaterial for the scattering coefficients due to the circular symmetry of the scatterer.

A critical point in Rayleigh's method is the fact that the fields can be equivalently written everywhere in the background medium using an alternative to Eq.(14) expansion, which follows from Green's theorem and expresses the total field as a superposition of outgoing cylindrical waves radiated by all scatterers in the lattice [43]

$$u(\vec{r}) = \sum_{\ell} \sum_{m=-\infty}^{+\infty} B_m^\ell \psi_m(\vec{r} - \vec{R}_\ell) \tag{15}$$

where the index $\ell$ runs over all cylinders – including $(0,0)$ – and $B_m^\ell = B_m e^{j\vec{K}\cdot\vec{R}_\ell}$ are the scattering coefficients of the wavefunctions radiated by cylinder $\ell$. By equating Eqs. (14) and (15), and using the addition theorem of cylindrical wavefunctions [44], we obtain a homogeneous system of the coefficients $B_m$ that allows the calculation of the Bloch wavevector $\vec{K}$:

$$\begin{aligned} & T_{|m|}^{-1} B_m - \sum_{n=-\infty}^{+\infty} S_{n-m} B_n = 0 \\ & \text{where} \quad S_{n-m} = \sum_{\ell \neq (0,0)} \psi_{n-m}(-\vec{R}_\ell) e^{j\vec{K}\cdot\vec{R}_\ell} \end{aligned} \tag{16}$$

Here $S_{n-m}$ is defined to be the lattice sum of order $n-m$. The latter expresses the coupling between wavefunctions of a given azimuthal order $m$ in the region of any scatterer to wavefunctions of order $n$ radiated by all the other scatterers in the lattice. In other words, the lattice sums represent lattice interaction constants between the harmonics that constitute the scattered field of the lattice elements. An efficient calculation method for the lattice sums is discussed in detail in Appendix A.

Equation (16) is the central result of Rayleigh's method and constitutes a homogeneous linear system of coefficients $B_m$, with the angular frequency $\omega$ and the Bloch momentum $\vec{K}$ acting as parameters. For compactness, the system can be expressed in matrix form as $\left(\mathbf{T}^{-1}-\mathbf{S}\right)\cdot\mathbf{b}=0$. Here $\mathbf{T}$ is the (diagonal) transmission matrix which only depends on the single rod geometry, $\mathbf{S}$ is the matrix of lattice sums which depends on the geometry of the lattice, and $\mathbf{b}$ is the vector of unknown scattered wave amplitudes. Hence the scatterers' properties and the geometry of the lattice appear separately in the final system through the matrices $\mathbf{T}$ and $\mathbf{S}$. The latter introduces a change in the topology felt by the scatterer and hence modifies the spectrum of the corresponding boundary value problem, as for example by shifting the existing resonances or introducing new ones [45, 46].

The homogeneous system (16) can be solved to evaluate the Bloch wavevector $\vec{K}$ and the scattering amplitudes $B_m$ up to a common factor that can be used to normalize the average energy of the Bloch mode. Since we are only interested in the electric dipole, magnetic dipole, and electric quadrupole contributions of the polarization current, the system can be truncated to the order $m=\pm 2$ resulting in a 5×5 system of equations. In the simpler TM case, for example, the system degenerates to a rank-2 3×3 system which yields for the first scattering coefficients

$$B_{-1}^{TM}=b_0\frac{S_2 D_0^{TM}-S_1^2}{S_1 D_{-1}^{TM}-S_{-1}S_2}, \quad B_0^{TM}=b_0, \quad B_1^{TM}=b_0\frac{S_{-1}S_1-D_{-1}^{TM}D_0^{TM}}{S_1 D_{-1}^{TM}-S_{-1}S_2} \qquad (17)$$

where $D_n^{TM}=S_0-1/T_{|n|}^{TM}$, the $S$ coefficients are the lattice sums defined in Eq. (16) and the value $b_0$ (in V/m) of the zero-order coefficient is arbitrary.

Once the scattering amplitudes $B_m$ are calculated, we need to relate them to the average volume polarization densities $\vec{P}_{av}, \vec{M}_{av}, \ddot{Q}_{av}$ of Eqs. (7) and (11). In our previous work [38] we have derived a multipole approximation of the EM field radiated by a single nonmagnetic cylinder keeping only the moments $\vec{p},\vec{m},\ddot{Q}$ of its polarization current:

$$\vec{E}^{scat}(\vec{r}) \cong Z_b \left( k_b^2 \vec{p} c_b G(\vec{r}) - 2jk_b \vec{m} \times \nabla G(\vec{r}) \right)$$

$$\vec{H}^{scat}(\vec{r}) \cong jk_b \vec{p} c_b \times \nabla G(\vec{r}) + k_b^2 \vec{m} G(\vec{r}) + \frac{j}{2} \nabla \times \left( \omega \ddot{Q} \cdot \nabla G(\vec{r}) \right)$$

(18)

where $k_b, c_b, Z_b$ are the wavevector, speed of light, and impedance in the background medium, and $G(\vec{r}) = (j/4) H_0^{(1)}(k_b |\vec{r}|)$ the scalar 2D Green's function. In the present context, where the cylinder is part of a lattice, the polarization current (and hence the moments) is simply the current developed in the cylinders as they scatter the field radiated by all others. Now, the multipole approximation of the fields in (18) imply truncated versions of the expansion (15). This is clear if one introduces into (18) the cylindrical wavefunctions of (13) (with the help of certain Bessel function recurrence formulas) and considers the longitudinal components. The result reads

$$E_z^{scat}(\vec{r}) \cong \frac{jZ_b k_b^2}{4} \left[ p_z c_b \psi_0(\vec{r}) + (m_x + jm_y) \psi_{-1}(\vec{r}) + (m_x - jm_y) \psi_1(\vec{r}) \right]$$

(19)

$$H_z^{scat}(\vec{r}) \cong \frac{jk_b^2}{8} \begin{bmatrix} 2m_z \psi_0(\vec{r}) - (p_x + jp_y) c_b \psi_{-1}(\vec{r}) - (p_x - jp_y) c_b \psi_1(\vec{r}) \\ + \frac{j\omega}{2} \left( Q_{xy} + j\frac{Q_{yy} - Q_{xx}}{2} \right) \psi_{-2}(\vec{r}) + \frac{j\omega}{2} \left( Q_{xy} - j\frac{Q_{yy} - Q_{xx}}{2} \right) \psi_2(\vec{r}) \end{bmatrix}$$

where the first equation refers to the TM and the second to the TE case. Connecting these to the expression (14) of Rayleigh's method allows us to obtain the following relations between the multipole moments and the scattering amplitudes

$$p_z c_b = \frac{4B_0^{TM}}{jZ_b k_b^2}, \qquad m_x = \frac{2}{jZ_b k_b^2} \left( B_1^{TM} + B_{-1}^{TM} \right), \qquad m_y = \frac{2}{Z_b k_b^2} \left( B_1^{TM} - B_{-1}^{TM} \right)$$

$$p_x = \frac{4j}{k_b^2} \left( B_1^{TE} + B_{-1}^{TE} \right), \qquad p_y = \frac{4}{k_b^2} \left( B_{-1}^{TE} - B_1^{TE} \right), \qquad m_z = \frac{4}{jk_b^2} B_0^{TE}$$

(20)

$$\omega Q_{xy} = -\frac{8}{k_b^2} \left( B_2^{TE} + B_{-2}^{TE} \right), \quad \omega (Q_{yy} - Q_{xx}) = \frac{16j}{k_b^2} \left( B_{-2}^{TE} - B_2^{TE} \right)$$

The average volume polarization densities can then be calculated from Eq. (5). Replacing (20) into (11) the effective permittivity and permeability tensors can be readily determined. In Appendix C we show that the resulting formulas, which are valid as long as $\vec{K}\cdot\vec{r}\ll 1$, degenerate to well-known quasistatic and static expressions in the respective quasistatic ($k_b r_c \to 0$) and static ($k_b \to 0$) limits.

### IV. NUMERICAL EXAMPLES FOR THE CASE OF CYLIDNRICAL SILICON RODS

In this section we present numerically calculated results of the theoretical model analyzed in the previous sections. For the effective medium properties of the lattice structure to be found from Eq. (11) for a given input frequency $\omega$, the corresponding Bloch wavevector value (dispersion relation) must be calculated for which the homogeneous linear system (16) (or its truncated version (C2) in Appendix C for the TM case) admits a non-trivial solution. For illustration purposes, we seek solutions along the edges of the irreducible ΓXM triangle of the first Brillouin zone, [47] i.e. Bloch modes with $0 \leq K_x \leq \pi/a$ and $K_y = 0$ (ΓX), $K_x = \pi/a$ and $0 \leq K_y \leq \pi/a$ (XM), and $0 \leq K_x = K_y \leq \pi/a$ (MΓ). We thus adopt a "fixed-frequency" approach for finding these Bloch modes which means that we fix the frequency ω and search for the roots in the interval where the momentum varies, e.g. in $0 \leq K_x \leq \pi/a$ in the case of $\vec{K}$ lying along the ΓX line. This process requires the calculation of the lattice sums of Eq. (16) over a dense grid of Bloch momentum values, which makes the absolutely (fast) convergent formula derived in the Appendix A particularly helpful. In order to demonstrate the effectiveness of our methodology, we compare these values with the ones obtained with the conventional quasistatic approximations for $k_b r_c \to 0$ [32, 48] (see also Appendix C) and, additionally, with the photonic bands obtained from the literature through a different full-numerical method.

We focus on a square lattice of lossless silicon cylindrical rods embedded in free space at optical frequencies. The rod radii is $r_c = 158\ nm$ and they are separated in the lattice by $a = b = 698\ nm$. We assume $\varepsilon_c = 12$ for the rod material. For simplicity, we examine TM modes only (propagating in-plane, i.e. $K_z = 0$) and include electric dipole and magnetic dipole moments, recalling that in the TM case the electric quadrupole moment reduces to a doubled magnetic dipole. In that case the determinant of the linear system (16) takes the form

$$|\mathbf{D}| \equiv \begin{vmatrix} S_0 - 1/T_1^{TM} & S_1 & S_2 \\ S_{-1} & S_0 - 1/T_0^{TM} & S_1 \\ S_{-2} & S_{-1} & S_0 - 1/T_1^{TM} \end{vmatrix} \quad (21)$$

If one uses the reflection symmetry of the lattice sums $S_{-m} = (-1)^m S_m$, the above determinant requires the computation of only 3 lattice sums per frequency and per Bloch wavevector $(K_x, K_y)$. Due to the truncation of the system and the finite sampling in the momentum space, the roots of the determinant appear as minima of its magnitude when the latter is plotted versus the independent variable, for example $|D|(K_x)$ along the ΓX edge or $|D|(K_y)$ along the XM edge. The roots are double-checked through the minima of the numerically calculated reciprocal condition estimator of the above matrix (Rcond) [49] as a function of the same variable. This criterion does not work the other way around as is obvious from the formula $\|D\| \cdot \|D^{-1}\|$ for the condition number. Hence not every minimum of the Rcond index corresponds to a root of the determinant. As realized from Eq. (A8) in the Appendix A, the lattice sums have poles at values of the Bloch momenta $\vec{K}$ that satisfy $|\vec{K} + \vec{K}_h| = k_b$, for some pair of integers pair $h = (h_1, h_2)$. At these points the corresponding entries of matrix D blow up and a dip appears in the Rcond graph. The situation is explained in Fig. 2. In (a), the magnitude of the zero-order lattice sum $S_0$ is plotted in a logarithmic scale versus the normalized Bloch momentum $K_x a/\pi$ and the normalized frequency $a/\lambda$. Note how the maxima of $|S_0|$ form a folded line, which is the light line $K_x = \omega/c$ (or $y = 0.5x$ in the normalized coordinates) folded into the first Brillouin zone. In (b) we fix the frequency at $a/\lambda = 0.4$ and plot the magnitudes of the determinant and the Rcond indicator of our matrix D. Notice that there are two minima of $|\text{Rcond}(D)|$. The one around $K_x a/\pi = 0.75$ corresponds to a minimum of the determinant and hence to a Bloch mode. The other corresponds to a maximum of $|D|$ and is at $K_x a/\pi = 0.8$, which lies exactly on the light line where the lattice sums blow up.

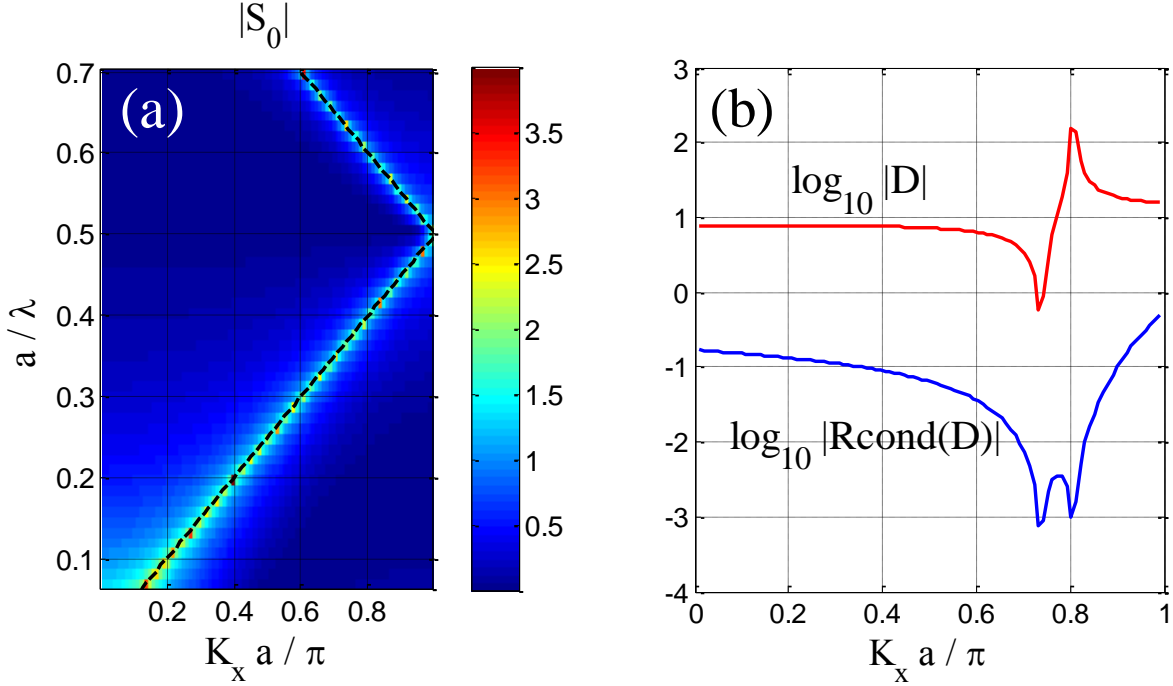

**Figure 2:** (a) Magnitude of the lattice sum $S_0$ in base-10 logarithmic scale, versus the normalized Bloch momentum $K_x a/\pi$ and the normalized frequency $a/\lambda$ for a lattice with $a = b = 698\,nm$ in air $(\varepsilon_b = 1)$. The dashed black line is the light line $K_x = \omega/c$ folded into the first Brillouin zone $K_x \leq \pi/a$. (b) Base-10 logarithm of the magnitude of the determinant (red curve) and of the inverse condition number (blue curve) of matrix D versus $K_x$ at $a/\lambda = 0.4$.

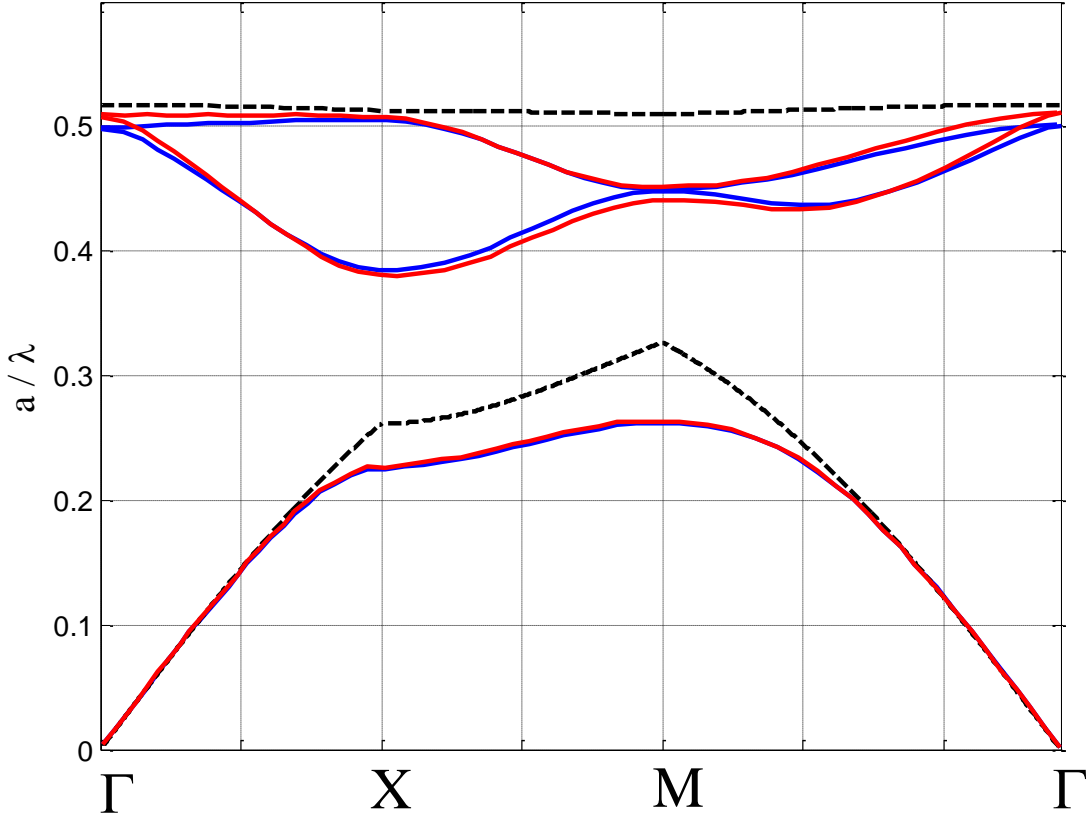

**Figure 3: TM photonic bands of a lattice of rods with $\varepsilon_c = 12$ and $r_c = 158\,nm$ with lattice constants $a = b = 698\,nm$ embedded in air $(\varepsilon_b = 1)$. The abscissa is Bloch momentum in $\pi/a$ units across the irreducible ΓXM triangle of the first Brillouin zone while the ordinate is the normalized frequency $a/\lambda$. The solid blue curves are the bands computed with a plane-wave expansion method obtained from work [22-25]. The red curves are the results of the present multipole method with a system truncated up to the magnetic dipole term. The dashed black curves are the predictions of the quasi-static approximation (C3).**

Figure 3 illustrates the results of our method in computing the bands of the metamaterial with the given parameters. The blue solid lines are the bands taken from work [22-25] for the same metamaterial using a plane-wave expansion method. This method is fully numerical without any underlying approximation for the wavelength or the size of the elements and limited only by the number of plane waves used to expand the field in the lattice and, accordingly, the two-dimensional Fourier transform of the periodic index modulation. Moreover, for TM problems in particular, it can be assumed to be offering a highly accurate estimation of the actual dispersion curves and it is usually considered a benchmark for assessing the performance of other methods, as for example here the quasistatic approximation or the truncated version of our multipole method. It should be emphasized however that, in the present case of cylindrical elements, our multipole method is also a highly accurate one and limited only by the number of cylindrical harmonics used to expand the field.

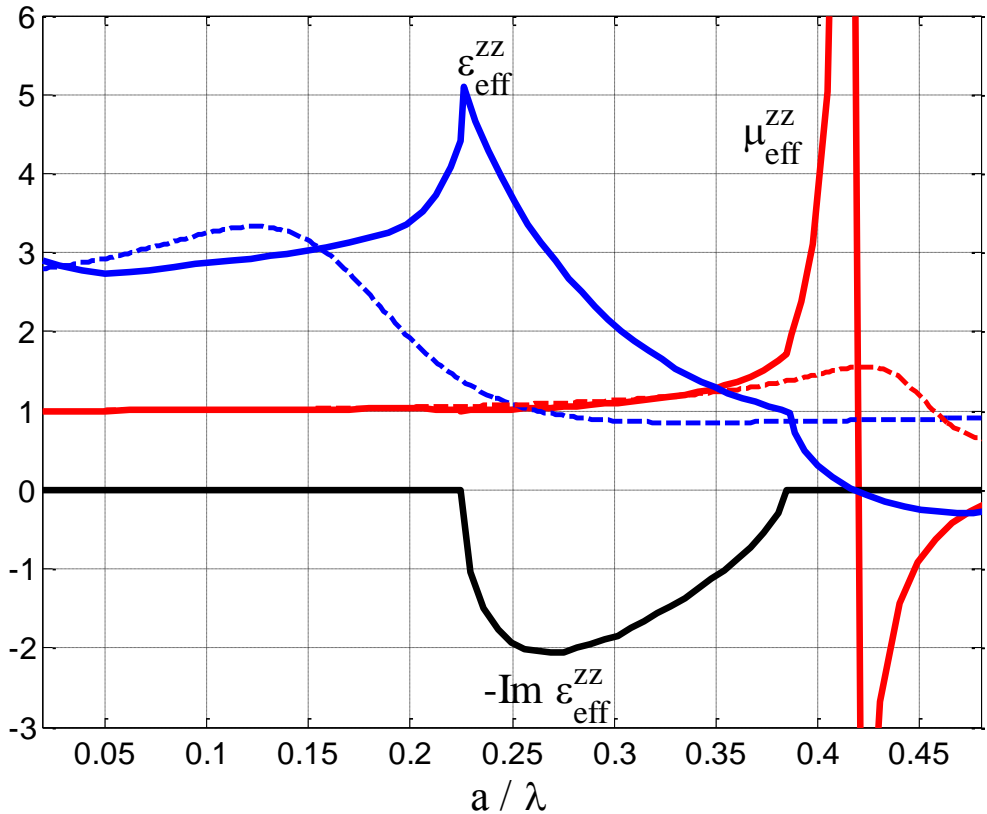

**Figure 4: Effective relative permittivity (blue) and permeability (red) with present method (solid lines) and with the quasi-static approximation (dashed lines). The black solid line is $-\mathrm{Im}(\varepsilon_{eff}^{zz})$ with the present method. The lattice parameters are the same with Fig. 3.**

This also implies greater computational efficiency compared to the more general-purpose plane-wave method since the cylindrical wave functions require truncation over a single integer variable, namely their order, while plane waves in two dimensions involve two continuous variables, namely $K_x, K_y$. What we will immediately see though and is remarkable is that our multipole method stays highly accurate even in its truncated version (21) which includes only the electric dipole and magnetic dipole moments. This is true over the range of frequencies that are free from higher-order resonances $(n \geq 2)$ which is usually adequate for exotic metamaterial properties to be manifested such as negative index and artificial magnetism. [22-25]

Also plotted in the Fig. 3 are the bands that we obtained with our method, namely from the singular points of the system (21). For illustration purposes we have used quite a fine frequency spacing in our solver so that even the very flat bands are reproduced, such as the top band along the ΓX edge. In all cases, and in the considered range of frequencies, the agreement between our method and the plane-wave expansion is indeed excellent. Notice how the lower band is almost perfectly reproduced by our solver. In the zero-frequency limit the dispersion curve becomes asymptotic to the straight line $K_x = 1.67\omega/c$, which implies a homogeneous effective dielectric medium with permittivity around 2.8, in excellent agreement with the Maxwell-Garnett mixing rule $\varepsilon_{eff}^{zz} = (1-f)\varepsilon_b + f\varepsilon_c = 2.76$ for a filling fraction $f = \pi r_c^2 / (ab) = 0.16$. The quasi-static approximation for the dispersion curves (C3) is also plotted in the same graph with a red curve and seen to fail to predict the actual bands, except in the very low frequency limit. Note for example the significant deviation of the quasi-static approximation near the X point of the ΓX edge or near the M point along the MΓ edge where it fails to predict the flattening of the band and hence the edge frequency of the bandgap.

Finally, in Figure 4 we plot the effective parameters $\varepsilon_{eff}^{zz}$ and $\mu_{eff}^{yy}$ obtained from Eqs. (11) and (B2) using the photonic bands that we have already determined in Figure 3. We restrict ourselves to Bloch waves propagating in the $x$ direction (ΓX edge of the irreducible triangle). For frequencies lying inside the bandgap we let our solver search for complex solutions. In this example, these frequencies are in the window $a/\lambda = 0.22 - 0.38$ and the corresponding Bloch wave vector is of the form $K_x = \frac{\pi}{a}(1+i\gamma)$. It is important to stress that the ability of our method to treat complex wave vectors lies in the alternative (reciprocal lattice) representation of the lattice sums that is derived in Appendix A and which can be analytically continued for complex $\vec{K}$, contrary to the direct summation in space (Eq. (16)). Superimposed in the same figure are also the quasi-static approximations of the effective parameters given by Eqs. (C6) and (C7), using the accurate expressions for the scattering coefficients. As expected already from the photonic band calculations, the two agree only in the low-frequency regime but not in higher frequencies where the quasi-static approximation fails to predict interesting regimes, as for example the negative-ε band starting around $a/\lambda = 0.415$. In this figure it is also interesting to note how the predictions of present method for the effective parameters are associated with the band structure of Fig. 3. The frequency window

$a/\lambda = 0.225 - 0.385$ corresponds to the first (lowest) bandgap and is characterized by a complex effective permittivity with a decreasing real part and a positive imaginary part that reaches a crest at $a/\lambda \sim 0.27$. The metamaterial therefore behaves effectively as a lossy or conductive medium inside the bandgap. At the upper edge of the bandgap ($a/\lambda \sim 0.385$), the permittivity becomes real again, still decreasing through positive values. The permittivity gets negative at $a/\lambda \sim 0.415$. At slightly higher frequencies the effective permeability becomes also negative having experienced a strong magnetic (dipole type) resonance. Hence for $a/\lambda \sim 0.42 - 0.50$ we have a double-negative (or left-handed) effective medium which is manifested in Fig. 3 as a photonic band with negative curvature (implying negative refraction) close to the Γ point.

## CONCLUSIONS

We presented a formulation for a dynamic theory to calculate the effective medium properties of 2D nonmagnetic metamaterial lattice structures. The effective medium properties are calculated as a function of the supported Bloch modes of the structure and of the multipole moments of the unit elements, which arise from the polarization currents excited. These in turn depend on the individual unit elements (material and geometry) as well as the lattice geometry. We propose techniques to evaluate the multipole moments, the Bloch modes and the resulting lattice sums. Finally, we present numerical examples that demonstrate the performance of the method compared to conventional quasistatic approaches. This is an extension of quasistatic effective medium formulas for 2D lattice structures.

## ACKNOWLEDGEMENT

This work was implemented within the framework of the Action "Supporting Postdoctoral Researchers" of the Operational Program "Education and Lifelong Learning" (Action's Beneficiary: General Secretariat for Research and Technology), and was co-financed by the European Social Fund and the Greek State (Program Nanokallos PE3_26). The authors would also like to thank the reviewers for constructive comments that helped to bring the paper to its final form.

## V.  APPENDICES

**APPENDIX A: Lattice sums**

The computation of the lattice sums $S_{n-m}$ is one of the most challenging tasks when utilizing Rayleigh's method. This is because the series of the definition (16) converges only conditionally which forbids a direct summation. To this end, sophisticated methods of computation must be employed. Here we follow the one presented in [50] which relies on the two equivalent representations of the quasi-periodic Green's function in the direct and reciprocal lattice spaces. The latter is defined as the solution of

$$\left(\nabla_{x,y}^2 + k_b^2\right)G_{\vec{K}}(\vec{r}) = -\sum_{\ell} e^{j\vec{K}\cdot\vec{R}_\ell}\delta(\vec{r} - \vec{R}_\ell) \tag{A1}$$

where $\ell$ runs over all lattice points including $(0,0)$. An obvious way to express $G$ is a superposition of phased Green's functions of the Helmholtz operator namely

$$G_{\vec{K}}(\vec{r}) = \frac{j}{4}\sum_{\ell} e^{j\vec{K}\cdot\vec{R}_\ell}\psi_0(\vec{r} - \vec{R}_\ell) \tag{A2}$$

This is the *direct lattice* representation. Separating the contribution of the source at $(0,0)$ and using the addition theorem of cylindrical wavefunctions to express terms $\ell \neq (0,0)$ in terms of nonsingular wavefunctions with argument $\vec{r}$, (A2) is written

$$G_{\vec{K}}(\vec{r}) = \frac{j}{4}\psi_0(\vec{r}) + \frac{j}{4}\sum_{m=-\infty}^{+\infty} S_{-m}\hat{\psi}_m(\vec{r}) \tag{A3}$$

which is valid for $|\vec{r}| < \min(|\vec{R}_\ell|) = \min(a,b)$, while $S_{-m}$ is the lattice sum defined in (16). An alternative way to express $G$ is obtained by noting that

$$\sum_{\ell} e^{j\vec{K}\cdot\vec{R}_\ell}\delta(\vec{r} - \vec{R}_\ell) = e^{j\vec{K}\cdot\vec{r}}\sum_{\ell}\delta(\vec{r} - \vec{R}_\ell) = \frac{1}{ab}\sum_{h} e^{j(\vec{K}+\vec{K}_h)\cdot\vec{r}} \tag{A4}$$

where the Fourier series representation of the 2D Dirac comb was used. The wavevectors of the Fourier series run over the *reciprocal lattice* as $\vec{K}_h = 2\pi\left(\hat{x}h_1 a^{-1} + \hat{y}h_2 b^{-1}\right)$, where $h = (h_1, h_2)$ is an arbitrary pair of integers. Substituting (A4) into (A1) we obtain

$$\left(\nabla_{x,y}^2 + k_b^2\right)G_{\vec{K}}(\vec{r}) = -\frac{1}{ab}\sum_{h} e^{j(\vec{K}+\vec{K}_h)\cdot\vec{r}} \tag{A5}$$

Now, it is easy to see that the solution of the above can be written as a plane wave series

$$G_{\vec{K}}(\vec{r}) = \frac{1}{ab} \sum_{h} \frac{e^{j(\vec{K}+\vec{K}_h)\cdot\vec{r}}}{|\vec{K}+\vec{K}_h|^2 - k_b^2}. \tag{A6}$$

The above is the *reciprocal lattice* representation of the quasi-periodic Green's function. Further, by using the Jacobi-Anger formula for transforming the plane waves into series of cylindrical wavefunctions, (A6) becomes

$$G_{\vec{K}}(\vec{r}) = \frac{1}{ab} \sum_{m=-\infty}^{+\infty} \left( j^m \sum_{h} \frac{J_m(|\vec{K}+\vec{K}_h|r) e^{-jm\arg(\vec{K}+\vec{K}_h)}}{|\vec{K}+\vec{K}_h|^2 - k_b^2} \right) e^{jm\theta} \tag{A7}$$

and *arg* is the argument (or angle) of a vector. Equations (A3) and (A7) are equivalent representations of the quasi-periodic Green's function and both have the form of a Fourier series in the azimuthal variable $\theta$. Invoking orthogonality, one obtains

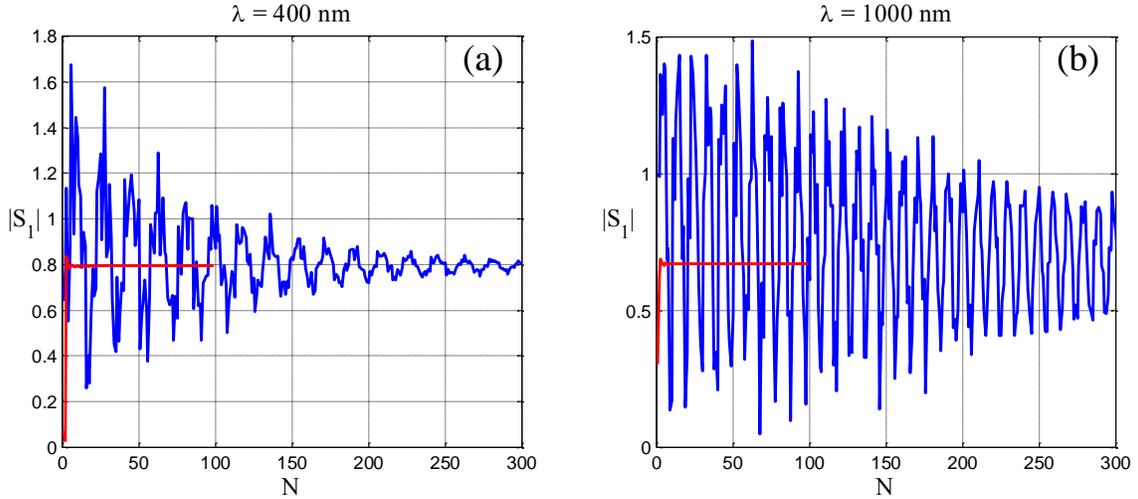

**Figure 5: Comparison of convergence between the direct (blue line) and reciprocal (red line) space formulas for lattice sums. Shown is the absolute value of the partial lattice sum $S_1^N$ versus the number of terms $N$ for a square lattice with $a = 698nm$, $K_x = 0.4\pi/a$, $K_y = 0$ at a wavelength (a) 400nm and (b) 1000nm. The background permittivity has been assumed $\varepsilon_b = 1 + j0.002$. In both cases the formula (A8) is terminated when a convergence of $10^{-6}$ is reached (in around 100 terms).**

$$S_{-m} J_m(k_b r) = -H_0^{(1)}(k_b r)\delta_{m,0} + \frac{4j^{m-1}}{ab}\sum_h \frac{J_m\left(\left|\vec{K}+\vec{K}_h\right|r\right)}{\left|\vec{K}+\vec{K}_h\right|^2 - k_b^2} e^{-jm\arg\left(\vec{K}+\vec{K}_h\right)}, \quad (A8)$$

$\delta_{m,0}$ being Kronecker's delta. Equation (A8) is an *absolutely convergent* expression for the lattice sums which converges at a rate $\left|\vec{K}_h + \vec{k}_0\right|^{-2.5}$ and is very effective for numerical calculations. This is the expression we use in this paper to compute the lattice sums that appear in the linear system (16). Moreover, if $N$ successive integrations over $r$ are applied to (A8), the speed of convergence can be improved to $\left|\vec{K}+\vec{K}_h\right|^{-N-2.5}$, which is another useful feature of this approach.

Closing this Appendix, it is interesting to actually visualize the superiority of the absolutely convergent expression for the lattice sums over the direct summation. This is shown in Fig. 5 at two different wavelengths, where a small imaginary part (loss) has been introduced into the background permittivity $\varepsilon_b$ in order to accelerate the convergence of the direct summation. Still, in the presence of loss, the summation over the reciprocal lattice needs less than 1/30 of the terms and less than 1/10 of the computation time to reach the same degree of convergence. Note also that formula (A8) converges fast and almost independent of the wavelength, contrary to the direct summation which slows down at lower frequencies. For a real $\varepsilon_b$, the direct summation cannot actually converge in a finite-precision machine.

**APPENDIX B: Effective parameters**

In this appendix we outline the derivation of the (*xx, yy, xy, yx*) effective parameters of the metamaterial lattice in the transverse direction. For a 2D configuration the tensors $\ddot{\varepsilon}_{eff}, \ddot{\mu}_{eff}$ of the relative effective permittivity and permeability have the form

$$\ddot{\varepsilon}_{eff} = \begin{pmatrix} \varepsilon_{eff}^{xx} & \varepsilon_{eff}^{xy} & 0 \\ \varepsilon_{eff}^{yx} & \varepsilon_{eff}^{yy} & 0 \\ 0 & 0 & \varepsilon_{eff}^{zz} \end{pmatrix}, \quad \ddot{\mu}_{eff} = \begin{pmatrix} \mu_{eff}^{xx} & \mu_{eff}^{xy} & 0 \\ \mu_{eff}^{yx} & \mu_{eff}^{yy} & 0 \\ 0 & 0 & \mu_{eff}^{zz} \end{pmatrix} \quad (B1)$$

where $\varepsilon_{eff}^{zz}$ and the 2×2 block of $\ddot{\mu}_{eff}$ take part only in the TM case, while $\mu_{eff}^{zz}$ and the 2×2 block of $\ddot{\varepsilon}_{eff}$ take part only in the TE case. In the former case, the second of (10) in combination with (9) imply the equations

$$H_{av,x} = \mu_{eff}^{xx}\left(H_{av,x} - 2M_{av,x}\right) + \mu_{eff}^{xy}\left(H_{av,y} - 2M_{av,y}\right)$$
$$H_{av,y} = \mu_{eff}^{yx}\left(H_{av,x} - 2M_{av,x}\right) + \mu_{eff}^{yy}\left(H_{av,y} - 2M_{av,y}\right)$$
(B2)

The average magnetic field and magnetic dipole moment are given by (7) and (20) with the scattering amplitudes resulting from the linear system (17). Splitting (B2) into their real and imaginary parts we get 4 equations for 8 unknowns, namely the real and imaginary parts of the permeability elements. This seemingly strange situation is resolved if one recalls that, for lossless media, $\ddot{\mu}_{eff}$ must be a Hermitian tensor, namely $\mu_{eff}^{xx}, \mu_{eff}^{yy}$ should be real and $\mu_{eff}^{xy} = \left(\mu_{eff}^{yx}\right)^*$, which reduces the number of unknowns to 4, i.e. as many as the equations.

The same procedure is followed in the TE case for the transverse elements of the permittivity tensor. The first of (10) implies

$$\varepsilon_0^{-1} P_{av,x} = \left(\varepsilon_{xx}^{eff} - \varepsilon_b\right) E_{eq,x} + \varepsilon_{xy}^{eff} E_{eq,y}$$
$$\varepsilon_0^{-1} P_{av,y} = \varepsilon_{yx}^{eff} E_{eq,x} + \left(\varepsilon_{yy}^{eff} - \varepsilon_b\right) E_{eq,y}$$
(B3)

The average electric dipole moments are given by (20) in terms of the scattering amplitudes. Regarding the electric field $\vec{E}_{eq}$ one first uses (7) and the first of (9) to find its expression in terms of the moments

$$\vec{E}_{eq} = \frac{\left(k_b^2 \ddot{I} - \vec{K}\vec{K}\right)\left(k_b \vec{P}_{av} c_b\right) + \frac{j\omega}{2}\vec{K}\times\left[\vec{K}\times\left(\ddot{Q}_{av}\cdot\vec{K}\right)\right] - k_b^2 \vec{K}\times\vec{M}_{av}}{\omega\varepsilon_0\varepsilon_b\left(|\vec{K}|^2 - k_b^2\right)}$$
(B4)

and then use (20).

**APPENDIX C: Static and quasi-static limits**

In this appendix we show that the dynamic effective medium formulas assume the correct values in the static and quasi-static limits. Only the TM case is presented; the TE case follows the same arguments. We begin with the lattice sums, as given in their reciprocal lattice representation by (A8), and consider their asymptotic expressions for $k_b \to 0$:

$$S_0 \sim -\frac{2j}{\pi}\ln(k_b r) - \frac{4j}{ab}\left(\left|\vec{K}\right|^2 - k_b^2\right)^{-1} \sim -\frac{4j}{ab}\left(\left|\vec{K}\right|^2 - k_b^2\right)^{-1},$$

$$S_{-m} \sim -\frac{4j}{ab} j^m \left(\frac{\left|\vec{K}\right|}{k_b}\right)^{|m|} \left(\left|\vec{K}\right|^2 - k_b^2\right)^{-1} e^{-jm\arg(\vec{K})} \sim S_0 j^m \left(\frac{\left|\vec{K}\right|}{k_b}\right)^{|m|} e^{-jm\arg(\vec{K})},$$

(C1)

The logarithmic term of $S_0$ can be neglected since (as is verified by the results), the difference $\left|\vec{K}\right|^2 - k_b^2$ is of commensurate order with $k_b^2$. This is suggested in (C1) by the second asymptotic expression for $S_0$. Thanks to this remark, the higher-order lattice sums can be expressed in terms of $S_0$ which facilitates the algebra. This is the second asymptotic expression for $S_{-m}$ in (C1).

The dispersion relation $\vec{K}(\omega)$ of the Bloch modes in the lattice follows from the roots of the determinant of the system (16), which in the TM case reads

$$\begin{vmatrix} S_0 - 1/T_1^{TM} & S_1 & S_2 \\ S_{-1} & S_0 - 1/T_0^{TM} & S_1 \\ S_{-2} & S_{-1} & S_0 - 1/T_1^{TM} \end{vmatrix} \equiv |\mathbf{D}| = 0 \qquad (C2)$$

When the approximations (C1) are used in (C2), the system matrix obtains the form $\mathbf{P}^{-1}\tilde{\mathbf{D}}\mathbf{P}$, where $\mathbf{P} = \text{diag}\left[1, \exp\left(j\arg(\vec{K})\right), \exp\left(2j\arg(\vec{K})\right)\right]$ is a diagonal unitary matrix that depends on the angle of the Bloch momentum $\vec{K}$ and $\tilde{\mathbf{D}}$ is a matrix which is obtained by setting $\arg(\vec{K}) = 0$ in the expressions of $S_{-m}$ in (C1). The conclusion is that the system's determinantal equation reduces to $|\tilde{\mathbf{D}}| = 0$ which depends only on the magnitude and not on the angle of $\vec{K}$. This is the mathematical proof of the expected fact that

the isofrequency curves of the lattice become circles in the limit $k_b \to 0$, where the wavelength is too long to resolve characteristic directions in the underlying lattice geometry. Solving $|\tilde{\mathbf{D}}| = 0$ for $|\vec{K}|$ yields

$$|\vec{K}|^2 \sim k_b^2 \left(1 + \tilde{T}_0\right) \frac{1 + \tilde{T}_1}{1 - \tilde{T}_1} \tag{C3}$$

where $\tilde{T}_{|m|} = -\frac{4j}{k_b^2 ab} T_{|m|}$ and the symbol of asymptotic equality reminds us that this implies the limit $k_b \to 0$.

Having determined the Bloch momentum in the long-wavelength regime, one can use it together with (C1) in (17) to compute the respective asymptotic expressions for the coefficients $B_{\pm 1,0}^{TM}$. The result is

$$B_{\pm 1,0}^{TM} \sim b_0 \frac{K_y \pm jK_x}{k_b} \times \frac{\tilde{T}_1}{\tilde{T}_0 \left(1 + \tilde{T}_1\right)} \tag{C4}$$

Subsequently, these can be plugged into (20) to obtain the electric and magnetic dipole moments

$$p_z c_b = \frac{4b_0}{jZ_b k_b^2}, \quad m_x = \frac{4K_y b_0}{jZ_b k_b^3} \times \frac{\tilde{T}_1}{\tilde{T}_0 \left(1 + \tilde{T}_1\right)}, \quad m_y = -\frac{4K_x b_0}{jZ_b k_b^3} \times \frac{\tilde{T}_1}{\tilde{T}_0 \left(1 + \tilde{T}_1\right)} \tag{C5}$$

which are finally substituted in the first of (11) to provide the effective permittivity in the long-wavelength limit

$$\varepsilon_{\text{eff}}^{zz} \sim \varepsilon_b \left(1 + \tilde{T}_0\right) \tag{C6}$$

The elements of the permeability tensor can be found by substituting into (B2) the asymptotic expressions of $\vec{M}_{av}$ and $\vec{H}_{av}$. The former is obtained from (C5) after dividing by $S_{cell} = ab$. The latter is simply given from (4) as $\vec{H}_{av} = \vec{K} \times \vec{E}_{av} / (\omega \mu_0)$, where from (9) and (10) (TM case) $\vec{E}_{av} = \hat{z} P_{av,z} / \varepsilon_0 \left(\varepsilon_{\text{eff}}^{zz} - \varepsilon_b\right)$. Then from equations (B2) we finally find

$$\mu_{eff}^{xy} \sim \mu_{eff}^{yx} \sim 0, \qquad \mu_{eff}^{xx} \sim \mu_{eff}^{yy} \sim \frac{1+\tilde{T}_1}{1-\tilde{T}_1} \tag{C7}$$

A first validation of our method comes from formulas (C3), (C6) and (C7) which combine to give $\left|\vec{K}\right|^2 \sim k_0^2 \varepsilon_{eff}^{zz} \mu_{eff}^{xx} \sim k_0^2 \varepsilon_{eff}^{zz} \mu_{eff}^{yy}$ as expected in the long-wavelength limit, where $k_0^2 = \omega^2 \mu_0 \varepsilon_0$ is the vacuum wavenumber. To further validation, we consider the special case where the rods are homogeneous with relative permittivity $\varepsilon_c$. From our previous work [38] we recall the long-wavelength formulas for the TM cylindrical scattering coefficients

$$T_0^{TM} \sim \frac{j\pi(k_b r_c)^2}{4} \frac{\left[2(\varepsilon_c/\varepsilon_b)-1\right]g_1(k_c r_c)-1}{g_1(k_c r_c)+1}$$

$$T_1^{TM} \sim \frac{j\pi(k_b r_c)^2}{4} \frac{g_1(k_c r_c)-1}{g_1(k_c r_c)+1} \tag{C8}$$

where $g_n(k_c r_c) = \frac{J_n(k_c r_c)}{(k_c r_c) J_n'(k_c r_c)}$. Substituting the above in (C6) and (C7) we get

$$\varepsilon_{eff}^{zz} \approx (1-f)\varepsilon_b + f\varepsilon_c \left[\frac{2g_1(k_c r_c)}{1+g_1(k_c r_c)}\right]$$

$$\mu_{eff}^{xx}, \mu_{eff}^{yy} \sim \left[1+f\frac{g_1(k_c r_c)-1}{g_1(k_c r_c)+1}\right] \times \left[1-f\frac{g_1(k_c r_c)-1}{g_1(k_c r_c)+1}\right]^{-1} \tag{C9}$$

where $f = \pi r_c^2 / ab$ is the filling fraction of the cylindrical inclusions in the lattice. These results agree exactly with previously published formulas for cylindrical inclusions in the quasi-static limit [28, 30, 32, 48], namely when $k_b r_c \ll 1$ but $k_c r_c$ stays finite, thus taking into account the resonances of the circular rods. If the wavelength is increased further to the static limit when $k_c r_c \to 0$ as well, then $g_1(k_c r_c) \to 1$, magnetic behavior vanishes while the first of (C9) reduces to the classic Maxwell-Garnett mixing formula.